# Phenotypic mixing and hiding may contribute to memory in viral quasispecies

Claus O Wilke*[1] and Isabel S Novella[2]

Address: [1]Digital Life Laboratory, California Institute of Technology, Mail Code 136-93, Pasadena, California 91125, USA and [2]Department of Microbiology and Immunology, Medical College of Ohio, Toledo, Ohio 43614, USA

Email: Claus O Wilke* - wilke@caltech.edu; Isabel S Novella - isabel@mco.edu

* Corresponding author

## Abstract

**Background:** In a number of recent experiments with food-and-mouth disease virus, a deleterious mutant, RED, was found to avoid extinction and remain in the population for long periods of time. Since RED characterizes the past evolutionary history of the population, this observation was called quasispecies memory. While the quasispecies theory predicts the existence of these memory genomes, there is a disagreement between the expected and observed mutant frequency values. Therefore, the origin of quasispecies memory is not fully understood.

**Results:** We propose and analyze a simple model of complementation between the wild type virus and a mutant that has an impaired ability of cell entry, the likely cause of fitness differences between wild type and RED mutants. The mutant will go extinct unless it is recreated from the wild type through mutations. However, under phenotypic mixing-and-hiding as a mechanism of complementation, the time to extinction in the absence of mutations increases with increasing multiplicity of infection (m.o.i.). If the RED mutant is constantly recreated by mutations, then its frequency at equilibrium under selection-mutation balance also increases with increasing m.o.i. At high m.o.i., a large fraction of mutant genomes are encapsidated with wild-type protein, which enables them to infect cells as efficiently as the wild type virions, and thus increases their fitness to the wild-type level. Moreover, even at low m.o.i. the equilibrium frequency of the mutant is higher than predicted by the standard quasispecies model, because a fraction of mutant virions generated from wild-type parents will also be encapsidated by wild-type protein.

**Conclusions:** Our model predicts that phenotypic hiding will strongly influence the population dynamics of viruses, particularly at high m.o.i., and will also have important effects on the mutation-selection balance at low m.o.i. The delay in mutant extinction and increase in mutant frequencies at equilibrium may, at least in part, explain memory in quasispecies populations.

## Background

RNA viruses are important pathogens in humans, animals, and plants. Their short generation time and high mutation rates allow them to adapt rapidly to changing environmental conditions, and make them ever-changing targets for anti-viral therapies or vaccinations. RNA viruses form highly polymorphic clouds of mutants, so-called quasispecies [1–3]. At viral loads of $10^8$ or more virions in an infected host, viral quasispecies *in vivo* often contain all possible single-point mutants from the consensus sequence, as well as a sizeable fraction of two- or three-point mutants [4]. Besides their importance as pathogens,

RNA viruses have in recent years also become one of the main tools for experimental verification of theoretical population genetics and evolutionary theory in general (reviewed in [3]). This line of research can draw on a substantial body of literature on the theory of quasispecies dynamics and its relation to standard population genetics [5–16].

The quasispecies theory has been useful to understand RNA virus evolution, but the assumptions underlying the model do not always reflect the biology of RNA viruses. In some instances the original model has been corrected to account for discrepancies, such as infinite vs. finite population sizes [7,9,16]. However, other factors need to be considered as well. For instance, the standard quasispecies model assumes low multiplicity of infection (m.o.i.), that is, each cell is assumed to be infected by at most one virion. At high m.o.i., when several virions coinfect a single cell, the virions can recombine [17–19] or reassort (for segmented viruses) [20–25]. Second, the model assumes that the genotype of a virion fully determines its fitness. However, this assumption is not always justified, since many viral functions can be provided in trans – an effect referred to as complementation. Thus, the phenotype of a virion may not reflect its genotype [26–29]. Phenotypic mixing and hiding is a particular case of complementation involving surface proteins (above citations). It occurs either at high m.o.i., when two or more different mutants or virus strains freely exchange genetic sequences and surface proteins inside a cell, or at low m.o.i., when a mutant offspring is encapsidated by its parent's proteins.

No general theory is available for the evolutionary dynamic of a viral quasispecies under the influence of virus density as it changes with changing m.o.i. However, some particular cases have been studied [30–32], mostly related to recombination and reassortment [33–36] and to the dynamic of defective interfering particles [37–41]. In this paper, we modify the quasispecies model by incorporating phenotypic hiding between wild-type virus and a deleterious mutant. We assume that the mutant differs from the wild type in a surface protein necessary for cell entry, so that the fitness difference between two virions is solely determined by the surface proteins, and not by the particular genomes they carry. We show that phenotypic hiding of the mutant genome behind wild-type capsids increases the mutant frequency above the level predicted by the standard quasispecies model. This increase in mutant frequency is most pronounced at high m.o.i., but exists also at very low m.o.i. Moreover, we show that the decrease of mutant frequency from an initially high level down to the level of mutation-selection balance is slowed down substantially at high m.o.i. because of phenotypic hiding.

We propose that our model may explain some of the quantitative differences observed between theoretical predictions of the quasispecies model and experimental data published in the literature [42–45]. Foot-and-mouth disease virus (FMDV) has been one of the most productive models of viral quasispecies dynamics (reviewed in [46]). FMDV belongs to the Picornaviridea family. It has a non-segmented, positive strand genome of approximately 8.5 Kb encapsidated by viral proteins, VP1, VP2, VP3 and VP4. A conserved Arginine-Glycine-Aspartic acid (RGD) triplet in βG-βH loop of VP1 recognizes the αvβ3, αvβ6 or αvβ1 integrin as viral receptor (reviewed in Ref. [47]), the first necessary step for virus entry. Work carried out with different RGD virus mutants and mimic peptides [48–52] clearly demonstrates the critical role that the RGD triplet has in virus entry, and its little tolerance to variation. Among the few RGD mutants that have been isolated, on is RED, a low fitness mutant in which the glycine is replace by glutamic acid [53]. Ruíz-Jarabo and co-workers found that after revertant RGD genomes became dominant, and mutant genomes were no longer detected in the consensus sequence, RED was able to remain in the population for very long times and at surprisingly high frequencies [42–45]. There is a qualitative agreement between the predictions of the quasispecies model and this result, but there is also a quantitative discrepancy. RED memory genomes can be observed at much higher frequency than expected. However, phenotypic hiding promotes a decelerated decay of mutant genomes in the population, and this decelerated decay could be interpreted as a memory effect, in agreement with the results by Ruíz-Jarabo et al. The research design of the original work [42–45] does not allow us to conclusively accept or reject our model as the correct explanation for their observations, but the values of mutant frequencies predicted by our model are consistent with the experimental data.

**Table 1: Mixing parameter *r* as a function of the m.o.i.**

| m.o.i. | r | m.o.i. | r | m.o.i. | r |
| --- | --- | --- | --- | --- | --- |
| 0.01 | 0.002 | 1.0 | 0.233 | 20 | 0.947 |
| 0.1 | 0.025 | 2.0 | 0.423 | 30 | 0.965 |
| 0.2 | 0.049 | 3.0 | 0.567 | 40 | 0.974 |
| 0.3 | 0.074 | 4.0 | 0.670 | 50 | 0.980 |
| 0.4 | 0.098 | 5.0 | 0.742 | 60 | 0.983 |
| 0.5 | 0.121 | 6.0 | 0.792 | 70 | 0.986 |
| 0.6 | 0.144 | 7.0 | 0.828 | 80 | 0.987 |
| 0.7 | 0.167 | 8.0 | 0.853 | 90 | 0.989 |
| 0.8 | 0.189 | 9.0 | 0.872 | 100 | 0.990 |
| 0.9 | 0.233 | 10.0 | 0.887 | 200 | 0.995 |

## Model

We assume that the virus is propagated under serial transfer, and model the change in mutant concentrations from passage to passage. We assume that the difference between wild type and mutant lies in the capsid proteins, such that the mutation affects the ability of the virus to enter a cell, but not the viability of the virus inside the cell. Under phenotypic mixing and hiding, there are four types of virions that can occur: wild-type genotype with wild-type phenotype, wild-type genotype with mutant phenotype, mutant genotype with wild-type phenotype, and mutant genotype with mutant phenotype. We will call virions whose genotype coincides with their phenotype as pure, and virions for which genotype and phenotype do not coincide as mixed. In the following paragraphs, we outline the derivation of a set of deterministic equations that describe the change in relative concentrations of the different pure and mixed virions in an infinite virus population. More details are given in the Methods section. Our model is similar to models studied by Szathmáry [38,39].

The probability $p_{ww}$ that an offspring virion is pure wild type is $p_{ww} = (1 - r)x_w + r x_w^2$ [Methods section, Eq. (14)], where $x_w$ is the probability with which a particular virion that infects a cell contains the wild-type genome, and $r$ is a parameter related to the multiplicity of infection. Likewise, the probabilities $p_{wm}$ and $p_{mw}$ that an offspring virion is mixed of either type are $p_{wm} = p_{mw} = rx_w x_m$ [Eq. (15)], where $x_m = 1 - x_w$, and the probability $p_{mm}$ that an offspring virion is pure mutant is $p_{mm} = (1 - r)x_m + r x_m^2$. The formula that relates the parameter $r$ to the multiplicity of infection, Eq. (13), is somewhat unwieldy. For reference, and in order to provide an intuitive meaning for $r$, we have tabulated $r$ (which we will call in the following also the mixing parameter) versus the m.o.i. in Table 1.

The probabilities $x_w$ and $x_m$ depend on the composition of the types of free virions on infection. Assume that $\gamma_{ww}$ is the concentration of pure wild-type virions at passage $t$, $\gamma_{wm}$ that of mixed virions with wild-type genotype and mutant phenotype, $\gamma_{mw}$ that of mixed virions with mutant genotype and wild-type phenotype, and $\gamma_{mm}$ that of pure mutant virions. Further, assume that virions with wild-type phenotype infect cells at a rate $\gamma_w$, while virions with mutant phenotype infect at a rate $\gamma_m < \gamma_w$. Then, the total rate at which cells are infected with wild-type genomes is $\gamma_w \gamma_{ww} + \gamma_m \gamma_{wm}$. Likewise, the total rate at which cells are infected with mutant genomes is $\gamma_w \gamma_{mw} + \gamma_m \gamma_{mm}$. Therefore, we have $x_w = (\gamma_w \gamma_{ww} + \gamma_m \gamma_{wm})/[\gamma_m(\gamma_{wm} + \gamma_{mm})]$. The probability $x_m$ follows from $x_m = 1 - x_w$.

Using the various formulas given in the previous two paragraphs, we find that the virion concentrations at passage $t + 1$ (primed quantities) are given by

$$\gamma'_{ww} = (1-r)x_w + rx_w^2, \quad (1a)$$
$$\gamma'_{wm} = rx_w x_m, \quad (1b)$$
$$\gamma'_{mw} = rx_w x_m, \quad (1c)$$
$$\gamma'_{mm} = (1-r)x_m + rx_m^2. \quad (1d)$$

We now introduce mutations into this model. We assume that the wild type mutates into the mutant with probability $\mu$. For simplicity, we assume that back mutations occur at the same rate. (It is safe to make this assumption, because the rate of back mutations from deleterious to advantageous mutants tends to have little effect on mutation-selection balance.) Mutations will typically not affect the phenotype of the offspring virion, which is determined by the distribution of mutant or wild-type capsid proteins in the cell. The extended model with mutations therefore becomes

$$\gamma'_{ww} = (1-\mu)\left[(1-r)x_w + rx_w^2\right] + \mu r x_w x_m, \quad (2a)$$
$$\gamma'_{wm} = (1-\mu)rx_w x_m + \mu\left[(1-r)x_m + rx_m^2\right], \quad (2b)$$
$$\gamma'_{mw} = (1-\mu)rx_w x_m + \mu\left[(1-r)x_w + rx_w^2\right], \quad (2c)$$
$$\gamma'_{mm} = (1-\mu)\left[(1-r)x_m + rx_m^2\right] + \mu r x_w x_m. \quad (2d)$$

## Results

### Extinction of mutant for $\mu = 0$

In order to evaluate whether the mutant can coexist with the wild type or will ultimately go extinct if it is not constantly regenerated through mutations, we have to analyze the fixed points of Eqs. (1a)–(1d). There are two trivial fixed points, corresponding to the cases in which all virions are either pure wild-type or pure mutant. We could not find any non-trivial fixed points. Since the expressions involved are extremely unwieldy, we searched for non-trivial fixed points only numerically (using the function NSolve of Mathematica [54]).

A linear expansion around the fixed points tells us whether the mutant can invade an established wild-type population or vice versa. If we introduce a small amount $\varepsilon$ of mutant virions into an established wild-type population, the total amounts of wild-type genomes $Y_w = \gamma_{ww} + \gamma_{wm}$ and mutant genomes $Y_m = \gamma_{mm} + \gamma_{mw}$ after the next passage become to first order in $\varepsilon$:

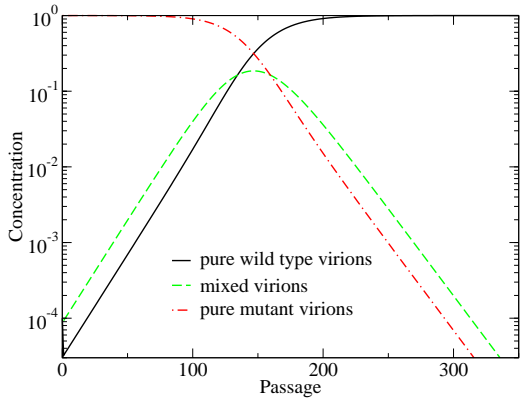

**Figure 1**
Invasion of an established mutant population by the wild type (m.o.i. = 5, $\gamma_m/\gamma_w$ = 0.8, $\mu$ = 0). Concentrations at passage 0 were $y_{ww}$ = 0.0001, $y_{mm}$ = 0.9999, $y_{wm} = y_{mw}$ = 0.

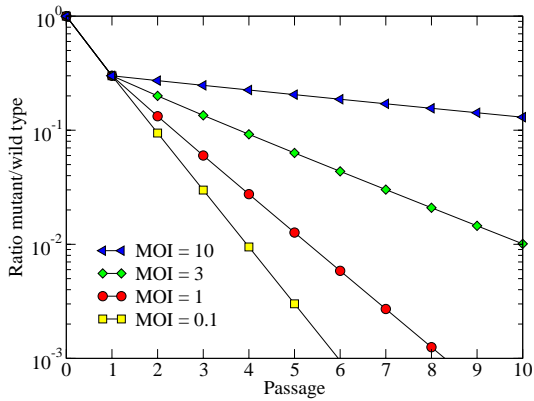

**Figure 2**
Ratio $Y_m/Y_w$ of mutant to wild type as a function of passage for various m.o.i. ($\gamma_m/\gamma_w$ = 0.3, $\mu$ = 0). Concentrations at passage 0 were $y_{ww} = y_{mm}$ = 0.5, $y_{mm} = y_{mw}$ = 0.

$$Y'_w = 1 - \gamma_m/\gamma_w \varepsilon, \quad (3a)$$
$$Y'_m = \gamma_m/\gamma_w \varepsilon. \quad (3b)$$

As long as $\gamma_m/\gamma_w < 1$ (as we assume throughout this paper), the amount of mutant genomes in the population decreases in the first passage, which implies that the mutant cannot invade. Likewise, if we introduce a small amount $\varepsilon$ of wild-type virions into an established mutant population, the total amounts of mutant and wild-type genomes after the next passage become to first order in $\varepsilon$:

$$Y'_w = \gamma_w/\gamma_m \varepsilon, \quad (4a)$$
$$Y'_m = 1 - \gamma_w/\gamma_m \varepsilon. \quad (4b)$$

In this case, the amount of the invading wild-type genomes in the population increases, which implies that the wild-type can invade an established mutant population. The invasion of a mutant population by a small amount of wild-type virions is shown in Fig. 1.

The preceeding analysis shows that Eqs. (1a)–(1d) have only a single stable fixed point, the one in which the whole population consists of pure wild-type virions. The mutant will therefore always go extinct eventually.

The rate at which the mutant disappears depends on the m.o.i. Figure 2 shows the ratio between mutant and wild-type concentrations as a function of passage, for various m.o.i. Concentrations are measured with respect to the genotype, that is, mutant and wild-type concentrations are $Y_m$ and $Y_w$, respectively. In the example given, we have $\gamma_m/\gamma_w$ = 0.3. At low m.o.i., the mutant-to-wild-type ratio drops from 1 at passage zero to values of $10^{-2}$ and lower after four passages. At high m.o.i., the reduction in mutant concentration proceeds much slower. For example, at an m.o.i. of three, the mutant-to-wild-type ratio is only $10^{-1}$ after four passages. An interesting aspect of Fig. 2 is that the reduction in mutant frequency during the first passage is independent of the m.o.i. This effect is a consequence of our assumption that the mutant has impaired ability of cell entry. In the first round of infection, when all virions are pure wild type or pure mutant, the mutant virions cannot benefit from phenotypic hiding. Therefore, their frequency will decrease corresponding to the true fitness difference to the wild type, independent of the m.o.i. From the second passage onward, the balance between pure and mixed virions is established, and further reduction of the mutant frequency is slowed down at high m.o.i.

Figure 3 shows the individual concentrations of the various pure and mixed virions at an m.o.i. of five. We observe that the concentration of pure mutant virions decreases much faster than that of mixed virions with mutant genomes. After ten passages both concentrations are equal, and after fifty passages, there are more than twice as many mixed virions with mutant genomes than there are pure mutant virions. When the mutant is rare, it is mostly hiding behind wild-type capsids, which slows down its extinction. Interestingly, the fraction of mutant genomes encapsidated with wild-type protein is equal to the fraction of wild-type genomes encapsidated with mutant pro-tein at zero mutation rate. This symmetry is a consequence of the symmetry of equations (1b) and (1c), and disappears for $\mu > 0$.

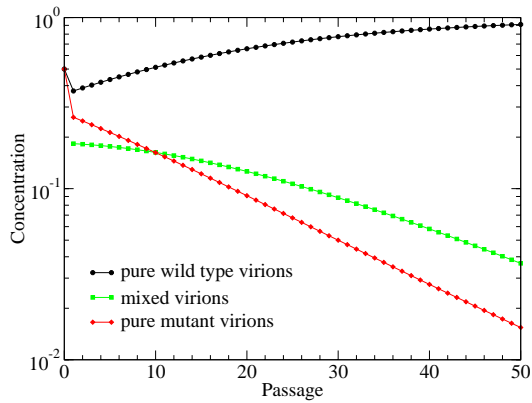

**Figure 3**
Relative concentrations of pure and mixed virions at m.o.i. = 5 ($\gamma_m/\gamma_w$ = 0.8, $\mu$ = 0). Concentrations at passage 0 were $y_{ww}$ = $y_{mm}$ = 0.5, $y_{mm}$ = $y_{mw}$ = 0.

*Mutation-selection balance for positive $\mu$*
We now turn to the general case of a positive mutation rate $\mu$. We first consider the case of vanishingly low m.o.i., $r = 0$. When we set $r = 0$, Eq. (2) becomes the discrete-time version of a model of maternal effects in quasispecies evolution [55]. In Ref. [55], the fitness of a virion is assumed to be the product of a contribution from the genotype of the virion's parent and one from the virion's own genotype. The situation here is simpler, because we assume that fitness is only determined by the phenotype (efficiency of cell entry). In the language of Ref. [55], this means we have $b_i = 1$ for all genotypes $i$. From Eqs. (5) and (6) in Ref. [55], we obtain the steady-state concentrations of pure and mixed virions (neglecting back mutations from mutant to wild type):

$$\gamma_{ww} = \frac{(1-\mu)(1-\mu-\gamma_m/\gamma_w)}{1-\gamma_m/\gamma_w}, \quad (5a)$$

$$\gamma_{wm} = \frac{\mu^2}{1-\gamma_m/\gamma_w}, \quad (5b)$$

$$\gamma_{mw} = \frac{\mu(1-\mu-\gamma_m/\gamma_w)}{1-\gamma_m/\gamma_w}, \quad (5c)$$

$$\gamma_{mm} = \frac{\mu(1-\mu)}{1-\gamma_m/\gamma_w}. \quad (5d)$$

We mentioned in the previous section that $\gamma_{wm} = \gamma_{mw}$ for $\mu = 0$. Clearly, this symmetry does not exist for $\mu > 0$. For $0 < \mu \ll 1$, the ratio in concentration of the two classes of mixed virions is $\gamma_{mw}/\gamma_{wm} = (1 - \mu - \gamma_m/\gamma_w)/\mu \approx 1/\mu$. There will typically be several orders of magnitude more mutant genomes encapsidated in wild-type protein than wild-type genomes encapsidated in mutant protein. Note, however, that this result holds only under the assumption of vanishingly low m.o.i. As the m.o.i. increases, the mixed virions with mutant genomes will still be at higher concentration than the mixed virions with wild-type genomes, but the ratio between the two will be much smaller than $1/\mu$. The ratio of mixed virions with mutant genomes to pure mutant virions is $\gamma_{mw}/\gamma_{mm} \approx 1 - \gamma_m/\gamma_w < 1$ at low m.o.i., that is, there are more pure virions than mixed virions with mutant genomes in equilibrium. We found numerically that at high m.o.i., this relationship is reversed, and there are substantially more mixed than pure mutants.

For the total amount of mutant genomes in the population, we obtain after summing Eqs. (5c) and (5d):

$$Y_m = \frac{\mu}{1-\gamma_m/\gamma_w}\left[2(1-\mu)-\gamma_m/\gamma_w\right]. \quad (6)$$

For comparison, the standard quasispecies model (without phenotypic hiding) predicts:

$$Y_m = \frac{\mu}{1-\gamma_m/\gamma_w}. \quad (7)$$

This expression is equivalent to the one given by Ruíz-Jarabo et al. in their analysis of quasispecies memory [44]. Equations (6) and (7) differ only in the factor $2(1 - \mu) - \gamma_m/\gamma_w$. This factor is larger than one for small mutation rates, that is, as long as

$$\mu < (1 - \gamma_m/\gamma_w)/2. \quad (8)$$

If the mutant has a substantial fitness disadvantage in comparison to the wild type (say at least 10%), then $\mu$ will always be small enough to satisfy this condition. Therefore, phenotypic hiding will lead to an increase in mutant genomes even at low m.o.i. The increase is at most a factor of two (for $\mu \ll 1$ and $\gamma_m/\gamma_w \ll 1$) in comparison to the prediction of the standard quasispecies model.

For the case of positive mutation rate and high m.o.i., there is no analytic solution available, and we have to resort to numerical simulations. Figure 4 shows the decay of mutant genomes over time, starting from an initial state where mutant and wild type are present in equal concentrations. We have displayed trajectories for a wide range of

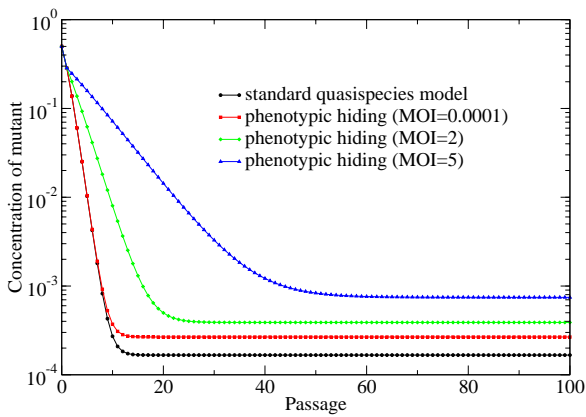

#### Figure 4
Concentration of mutant genomes $Y_m$ as a function of passage ($\gamma_m/\gamma_w = 0.4$, $\mu = 10^{-4}$). Concentrations at passage 0 were $y_{ww} = y_{mm} = 0.5$, $y_{wm} = y_{mw} = 0$. Trajectories labeled "phenotypic hiding" were generated from equations (2a)–(2d). The trajectory labeled "standard quasispecies model" was generated from iterating $Y'_m = [\gamma_m (1 - \mu) Y_m + \gamma_w \mu(1 - Y_m)]/[\gamma_m Y_m + \gamma_w(1 - Y_m)]$.

m.o.i., and we have also displayed the corresponding trajectory generated by the standard quasispecies model. All trajectories show the same qualitative behavior: The mutant frequency decays quickly during the initial passages, and then levels off in mutation-selection balance. As discussed in the previous paragraphs, the level of mutant genomes in mutation-selection balance is always higher in our model than in the standard quasispecies model. Moreover, the level of mutant genomes increases with increasing m.o.i. The initial decay of mutant genomes is also influenced by the m.o.i., and slows down with increasing m.o.i. In general, the initial decay of mutant genomes for $\mu > 0$ is identical to the one for $\mu = 0$, as described in the previous subsection. It is interesting to note that phenotypic hiding influences the equilibrium concentration of mutant genomes at low m.o.i., but not the speed at which equilibrium is attained. The initial decay of mutant genomes is identical for our model at low m.o.i. and for the standard quasispecies model.

### Discussion

Ruíz-Jarabo et al. measured the frequency of the minority FMDV RED mutant up to passage 50, and found that the frequency was still declining at passage 50 [44]. They observed mutant frequencies at passage 50 between $6.6 \times 10^{-4}$ and $10^{-3}$. These results are in agreement with our model of phenotypic hiding if we use a relative fitness of 0.4 for the RED mutant as compared to the wild type [42,44], and assume that (1) the fitness disadvantage is fully explained by impaired entry, (2) the mutation rate is $\mu = 10^{-4}$, and (3) the m.o.i. is approximately five (Fig. 4). The m.o.i. in Ref. [44] was between one and five. Ruíz-Jarabo et al. noted that mutant concentrations around $10^{-3}$ would follow from equilibrium state of the standard quasispecies model if the mutation rate were $\mu = 10^{-3}$ rather than $\mu = 10^{-4}$. However, the standard quasispecies model predicts that the mutation-selection balance should be reached around passage 10 (Fig. 4), which is at odds with the observation that the frequency of RED mutants continued to decline up to passage 50.

It is important to note that capsid proteins can be provided in trans during the production of infectious FMDV particles [51], and phenotypic hiding has been known for many years in a variety of picornaviruses [56–58]. Other factors may enhance or reduce the effects of complementation. For instance, compartimentalization or physical localization should prevent to some degree the free exchange of genomes and products.

In order to test fully the predictions of our model, we would need data on the decay of the mutant frequency from replicate populations propagated at different m.o.i. Unfortunately, such data does not exist. Although Ruíz-Jarabo et al. propagated their populations at an m.o.i. of 0.1 in their initial study [42] and at an m.o.i. between one and five in their follow-up study [44], we cannot compare the mutant frequencies reported in these two studies, because the method of screening for RED mutants was changed in the follow-up study ([44], Table 1, Note b). Moreover, a caveat applies to the comparison of the experimental data to our model: Ruíz-Jarabo et al. determined the frequency of RED mutants by first screening for monoclonal antibody (MAb SD6) resistant virions, and then determining the fraction of RED mutants among them from sequence data. While this method successfully prevents against counting MAb-resistant mutants that are not RED, it does not measure the fraction of RED mutants that hide behind RGD capsids. If phenotypic hiding plays an important role in this system, then the fraction of hiding RED mutants could be substantial, and the reported mutant frequencies would likely represent underestimates. In summary, the experimental data are consistent with our model, but do not allow us to either accept or reject it with certainty. We have shown in the present paper that phenotypic hiding can increase equilibrium frequencies by factors of two to five, or even higher, for high m.o.i. It is conceivable that other types of complementation can have similar effects [59].

Phenotypic hiding can bias fitness determinations if the phenotype, rather than the genotype, is quantified. This bias occurs simply because a fraction of mutant genomes will hide behind wild-type envelopes and will be missed, an effect that has been previously reported [29,60]. As a byproduct of our study, we predict two phases of mutant frequency decay specific to phenotypic hiding. In the first round of infection, all virions are pure wild type or pure mutant, so that the mutant virions cannot benefit from complementation. Therefore, their frequency will decrease corresponding to their true fitness and independently of the m.o.i. From the second passage onward, the balance between pure and mixed virions is established, and further reduction of the mutant frequency proceeds much slower if the m.o.i. is sufficiently high, as illustrated in Figure 2. As a consequence, if fitness competitions are halted after the first passage, measurements of the phenotype will not bias fitness determinations. These effect should not be observed in other forms of complementation, but should be characteristic of any situation in which the phenotype is determined by envelope proteins. Host range will often belong to this category, since receptor recognition is one of its main determinants. Other examples include antibody resistant mutants and escape mutants when the antiviral drug targets entry, such as rimantidine and amantidine-resistant influenza strains or WIN compound-resistant picornaviruses [61]. Thus, phenotypic hiding can keep a relatively large frequency of mutants in the absence of selective pressure that can provide a quick response when such selective pressure is restored.

Complementation can also complicate the interpretation of other experimental results involving competition between two (or more) strains. For instance, Miralles et al. [62,63] tested the effect of clonal interference in VSV populations by changing the population size in competitions between wild type and a MARM mutant. This MARM mutant carries a point mutation that confers resistance to I1 Mab, but it is neutral otherwise [64]. Clonal interference occurs in asexual organisms when population sizes are large enough to allow the simultaneous generation of more than one advantageous mutation, so that the best mutation has to compete not only with an inferior wild type but also with the other fitter mutants [65]. Thus, while the best possible mutation should always be the one becoming dominant, the rate at which it does so should be slowed down in large populations. The VSV competition experiments were carried out at increasing viral population sizes, but the number of infected cells was held constant, that is, the m.o.i. was increased alongside the population size. Therefore, complementation may also have contributed to the slowdown of the speed of adaptation at large population sizes, and the relative contributions of complementation and clonal interference to the slowdown would be hard to disentangle. The only phenotypic hiding effect that the authors could avoid were related to the determination of changes in MARM C frequencies, and these only correlate to the changes in beneficial mutation frequencies if the beneficial mutation mapped to the G glycoprotein.

Many infections in plants and animals are probably initiated by one or few viral particles, thus corresponding to a situation of low m.o.i. As replication proceeds and titers rise, the likelyhood of coinfection increases. The best evidence for multiple infections *in vivo* comes from the observation of natural recombinants. In the case of HIV-1, for instance, rampant recombination has been reported, even among different subtypes [66,67], and the number of proviruses per cell was estimated to be approximately three. Another well known example is influenza virus, for which reassortment is a major mechanism of evolutionary change [68]. As in the case of HIV, reassortment depends of coinfection, and once coinfection occurs, complementation is as likely to ensue as reassortment.

## Conclusions

We have shown that phenotypic hiding has substantial influence on the equilibrium frequencies of mutants in selection-mutation balance, and on the speed at which deleterious mutants are lost from the population. Our theory makes five predictions that can be tested experimentally:

1. The rate at which the mutant disappears should decrease and the equilibrium concentration of the mutant should increase with increasing m.o.i.

2. A substantial fraction of mutant genomes should be encapsidated with wild-type protein.

3. The ratio of mutant genomes encapsidated in wild-type protein to wild-type genomes encapsidated in mutant protein should decrease with increasing m.o.i.

4. Even at low m.o.i., the fraction of mutant present in mutation-selection balance should be larger (up to a factor of two) than predicted by the standard quasispecies model if the mutant fitness is low compared to the wild-type fitness.

5. The decrease of mutant frequency during the first passage should be independent of the m.o.i., and equal to that measured for later passages at very low m.o.i.

Future experiments can be designed to test these predictions, and can thereby clarify whether phenotypic hiding is a major contributor to viral memory and elevated equilibrium frequencies of mutants.

## Methods

In this section, we present some calculations that form the basis of the derivation of our model equations.

We first calculate the probability that a virion produced by an *n*-fold coinfected cell is pure. Assume a cell is infected by $k$ virions with wild-type genotype, and by $n - k$ virions with mutant genotype. Then, the probability that an offspring virion is pure wild-type is the probability that it has wild-type genotype, which is $k/n$, times the probability that it has wild-type phenotype, which is again $k/n$. Thus, the total probability is $(k/n)^2$. If we want to know the probability that an offspring virion is pure for a number of coinfected cells with different ratios in their multiplicity of infection of wild-type and mutant genotypes, then we have to sum over the different contributions of all the individual cells. In practice, we will not know exactly which cells are infected with what virions. However, it is reasonable to assume that we know the overall probabilities with which cells are infected with either mutant or wild-type genotype. Assume that $x_w$ is the probability that a cell is infected with wild-type genotype, and $x_m = 1 - x_w$ likewise for the mutant genotype. Then, the number of wild-type genomes $k$ in a cell is binomially distributed, and therefore the overall probability $\xi_{ww}(n)$ that an offspring virion from a cell infected by $n$ virions is pure wild-type becomes [38]

$$\xi_{ww}(n) = \sum_{k=1}^{n} \frac{k^2}{n^2} \binom{n}{k} x_w^k x_m^{n-k} = \frac{x_w x_m}{n} + x_w^2. \qquad (9)$$

Likewise, we find for the probability $\xi_{mm}(n)$ that an offspring virion is pure mutant $\xi_{mm}(n) = x_w x_m/n + x_m^2$.

Now we consider mixed virions. Assume again that a cell is infected by $k$ virions with wild-type genotype, and by $n - k$ virions with mutant genotype. In analogy to the considerations for a pure virion, we find that the probability of a mixed virion is $k(n - k)/n^2$, regardless of whether the genotype is wild-type and the phenotype is mutant or vice versa. The probabilities $\xi_{wm}(n)$ and $\xi_{mw}(n)$ that an offspring virion is mixed of either type are then

$$\xi_{wm}(n) = \xi_{mw}(n) = \sum_{k=1}^{n} \frac{k(n-k)}{n^2} \binom{n}{k} x_w^k x_m^{n-k} = \frac{n-1}{n} x_w x_m. \qquad (10)$$

Equations (9) and (10) are valid if all cells have exactly the same multiplicity of infection. However, infection is a Poisson process, and some cells will end up being infected by more virions than other cells. Assume that $\lambda$ is the multiplicity of infection of the whole population. Then, the probability that a single infected cell is infected by $k$ virions is $e^{-\lambda} \lambda^k / [(1 - e^{-\lambda}) k!]$. The factor $1 - e^{-\lambda}$ in the denominator corrects for the $e^{-\lambda}$ cells that will not be infected at all. We obtain the overall probabilities to find a pure or mixed offspring virion by summing over all possible combinations of infections, with their respective probabilities. For example:

$$p_{ww} = \frac{e^{-\lambda}}{1-e^{-\lambda}} \sum_{k=1}^{\infty} \frac{\lambda^k}{k!} \xi_{ww}(k) = \frac{e^{-\lambda}}{1-e^{-\lambda}} \sum_{k=1}^{\infty} \frac{\lambda^k}{k!} \left( \frac{x_w x_m}{k} + x_w^2 \right). \qquad (11)$$

After taking the sum, we find:

$$p_{ww} = x_w^2 + c x_w x_m, \qquad (12)$$

where the constant $c$ is given by

$$c = \frac{e^{-\lambda}}{1-e^{-\lambda}} \sum_{k=1}^{\infty} \frac{\lambda^k}{kk!} = \frac{e^{-\lambda}}{1-e^{-\lambda}} \left[ \mathrm{Ei}(\lambda) - \ln(\lambda) - \gamma \right]. \qquad (13)$$

Here, $\mathrm{Ei}(z)$ is the exponential integral, $\mathrm{Ei}(z) = -\int_{-z}^{\infty} \frac{e^{-t}}{t} dt$, and $\gamma$ is the Euler constant, $\gamma \approx 0.577216$. Finally, by making use of $x_w = 1 - x_m$, we arrive at

$$p_{mm} = c x_w + (1-c) x_w^2. \qquad (14)$$

Similarly, for a mixed virion, we find

$$p_{wm} = \frac{e^{-\lambda}}{1-e^{-\lambda}} \sum_{k=2}^{\infty} \frac{\lambda^k}{k!} \xi_{wm}(k) = \frac{e^{-\lambda}}{1-e^{-\lambda}} \sum_{k=2}^{\infty} \frac{\lambda^k}{k!} \frac{k-1}{k} x_w x_m = (1-c) x_w x_m. \qquad (15)$$

We obtain $p_{mm}$ and $p_{mw}$ from Eqs. (14) and (15) by interchanging w and m in all subscripts. Equations (14) and (15) show that the mixing parameter $r$ used in the main body of this paper is related to $c$ via $r = 1 - c$.

## Authors' contributions

COW was responsible for the analytic calculations. Both authors contributed substantially to the design of the study and the writing of the manuscript. Both authors read and approved the final manuscript.

## Acknowledgments

COW was supported by the NSF under Contract No. DEB-9981397; ISN was supported by NIH grant AI45686.